\documentclass{raa}            

\usepackage{graphicx,times}             
\usepackage{natbib}
\usepackage{amssymb,amsmath}
\bibpunct{(}{)}{;}{a}{}{,}

\usepackage{xcolor}
\usepackage{multirow}
\usepackage[T1]{fontenc}
\usepackage{ae,aecompl}
\usepackage{newtxtext,newtxmath}

\def\o5007{[O~{\sc iii}]$\lambda5007$\AA\ }
\def\f3e{[O~{\sc iii}]$\lambda5007$\AA(E)}
\def\oiiin{[O~{\sc iii}]$\lambda5007$\AA(N)}

\def\n6583{[N~{\sc ii}]$\lambda6583$\AA\ }
\def\sii{[S~{\sc ii}]$\lambda6717$\AA\ }
\def\siii{[S~{\sc ii}]$\lambda6731$\AA\ }

\def\a6300{[O~{\sc i}]$\lambda6300$\AA\ }
\def\b6363{[O~{\sc i}]$\lambda6363$\AA\ }

\def\obj{SDSS J1042-0018}

\usepackage[pagebackref=true]{hyperref}

\begin{document}

\title{SDSS J1042-0018 a broad line AGN but mis-classified as a HII galaxy in the BPT diagram 
by flux ratios of narrow emission lines}


   \volnopage{Vol.0 (20xx) No.0, 000--000}      
      \setcounter{page}{1}          

   \author{Cao Yi 
      \inst{1}
   \and Zhao SiDan 
      \inst{1}
   \and Zhu XingYu 
         \inst{1}
   \and Yu HaiChao 
         \inst{1}
   \and Wang YiWei 
         \inst{1}
   \and Zhang XueGuang$^{*}$ 
      \inst{1}
   }


\institute{School of Physics and technology, Nanjing Normal University, No. 1, Wenyuan Road,
	Nanjing, 210023, P. R. China; {\it{\color{blue}xgzhang@njnu.edu.cn}}\\
\vs\no
   {\small Received 20xx month day; accepted 20xx month day}}

\abstract{
In the manuscript, we discuss properties of the \obj~ which is a broad line AGN but mis-classified
as a HII galaxy in the BPT diagram (\obj~ called as a mis-classified broad line AGN). The emission 
lines around H$\alpha$ and around H$\beta$ are well described by different model functions, considering 
broad Balmer lines to be described by Gaussian or Lorentz functions. Different model functions lead 
to different determined narrow emission line fluxes, but the different narrow emission line flux 
ratios lead the \obj~ as a HII galaxy in the BPT diagram. In order to explain the unique properties 
of the mis-classified broad line AGN \obj, two methods are proposed, the starforming contributions 
and the compressed NLRs with high electron densities near to critical densities of forbidden emission 
lines. Fortunately, the strong starforming contributions can be preferred in the SDSS J1042-0018. 
The mis-classified broad line AGN \obj, well explained by starforming contributions, 
could provide further clues on the applications of BPT diagrams to the normal broad line AGN.
}
\keywords{galaxies: active --- galaxies: nuclei --- (galaxies:) quasars: emission lines --- galaxies: 
individual (SDSS J1042-0018)}

   \authorrunning{Cao et al.}            
   \titlerunning{mis-classified broad line AGN \obj}  

   \maketitle

%
%
\section{Introduction}           
\label{sect:intro}

\begin{figure}
\centering\includegraphics[width = \textwidth,height=0.6\textwidth]{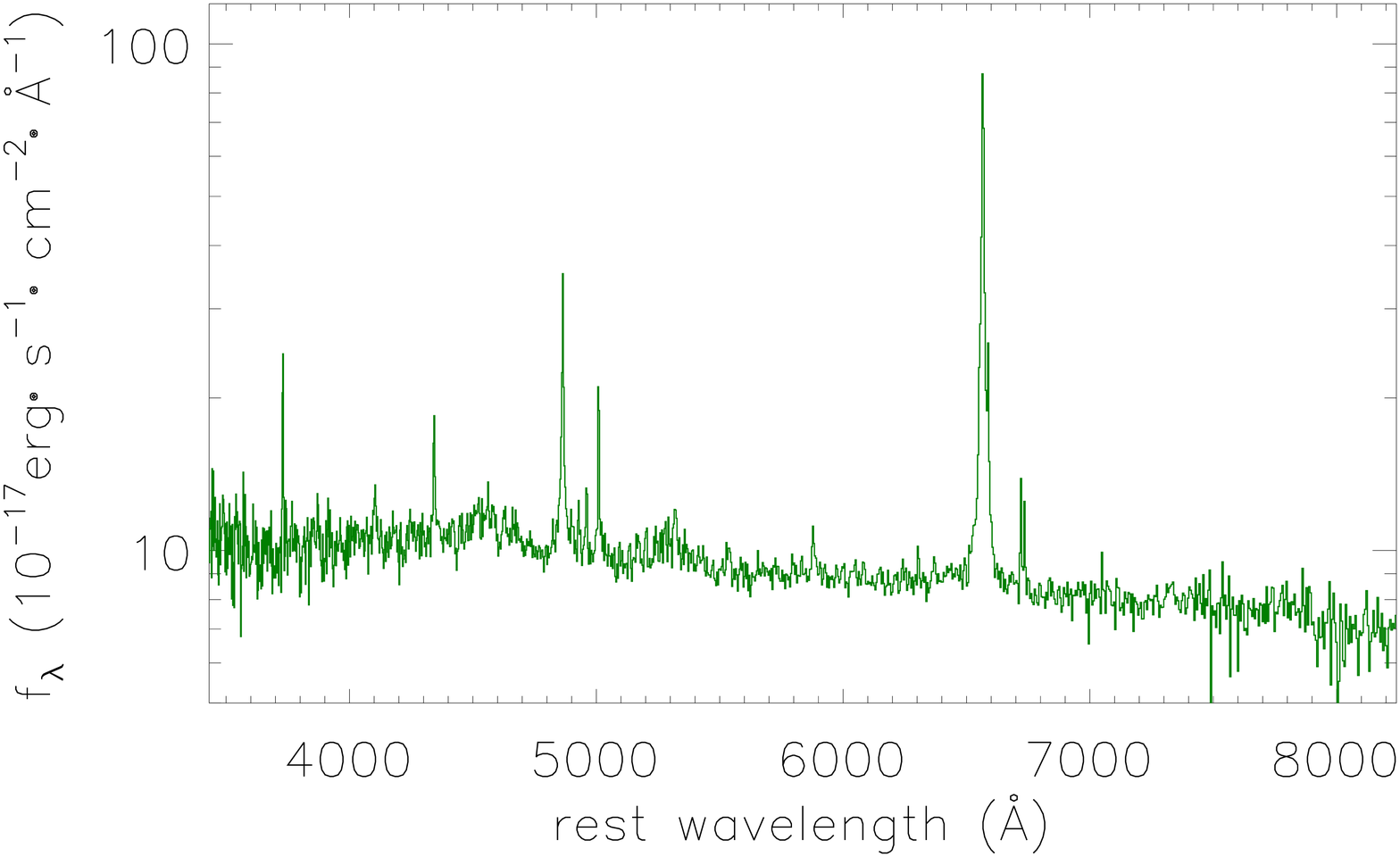}
\caption{The SDSS fiber spectrum of \obj.
}
\label{spec}
\end{figure}

	\obj~ (=SDSS J104210.03-001814.7) is a common broad line AGN with apparent broad 
emission lines, as the shown spectra in Fig.~\ref{spec} from SDSS (Sloan Digital Sky Survey). 
However, based on flux ratios of narrow emission lines well discussed in Section 2, \obj~ can 
be well classified as a HII galaxy in the BPT diagram \citep{bpt81, kb01, ka03, kb06, ke13a, 
kb19, zh20}. Therefore, in the manuscript, some special properties of \obj~ is studied and 
discussed.

	Under the commonly accepted and well-known constantly being revised Unified Model 
\citep{an93, ra11, nh15, aa17, bb18, bn19, kw21}, Type-1 AGN (broad line Active Galactic Nuclei) 
and Type-2 AGN (narrow line AGN) having the similar properties of intrinsic broad and narrow 
emission lines, however, Type-2 AGN have their central broad line regions (BLRs) with distances
of tens to hundreds of light-days \citep{kas00, kas05, bp06, bp09, dp10, ben13, fa17} to central
black holes (BHs) totally obscured by central dust torus (or other high density dust clouds),
leading to no broad emission lines (especially in optical band) in observed spectra of Type-2 AGN.
Meanwhile, Type-1 AGN and Type-2 AGN have the similar properties of narrow emission lines, due to 
much extended narrow emission line regions (NLRs) with distances of hundreds to thousands 
of pcs (parsecs) to central BHs \citep{za03, fi13, ha14, hb14, fi17, sg17, zh17}.

	For an emission line object, two main methods can be conveniently applied to classify 
whether the object is an AGN or not? On the one hand, an broad line object with clear spectroscopic
features of broad emission lines and the strong blue power-low continuum emissions can be directly 
classified as a type-1 AGN. On the other hand, for a narrow line object, the well-known BPT diagrams 
can be conveniently applied to determine which narrow line object is a Type-2 AGN (narrow line AGN) or
a HII galaxy by properties of flux ratios of narrow emission lines. Therefore, the flux ratios of 
narrow emission lines for broad line objects (type-1 AGN) can also lead to the objects well classified 
as AGNs in the BPT diagrams. However, there are some special bright type-1 AGNs (we called them as the 
mis-classified AGNs), of which flux ratios of narrow emission lines lead the AGNs lying in the regions 
for HII galaxies in the BPT diagrams. More recently, \citet{zh21x} have reported the well identified 
quasar SDSS J1451+2709 as a mis-classified quasar, due to its mis-classification as a HII galaxy in 
the BPT diagrams through narrow emission line flux ratios, after considering different model functions 
applied to describe the emission lines. In the manuscript, the second mis-classified broad line AGN 
\obj~ is reported and discussed.

	As discussed in \citet{zh21x}, Type-1 AGN have more complicated line profiles of emission lines 
which will be discussed by different model functions leading to quite different properties of narrow 
emission lines. As the shown results in SDSS J1451+2709\ in \citet{zh21x}, there are 
intemediate broad emission lines have line width (second moment) around several hundreds of kilometers 
per second, gently wider than extended components of narrow emission lines, such as the commonly known 
extended components in [O~{\sc iii}]$\lambda4950, 5007$\AA~ doublet discussed in \citet{gh05, sh11, 
zh17, zh21a, zh21m}. Once there are multi-epoch spectra, variability properties of emission components 
can be applied to determine whether one emission component is from NLRs. In the manuscript, although 
there is only single-epoch spectrum of \obj, flux ratios of narrow emission lines can be applied to 
determine the classifications of \obj~ in the BPT diagram. In the manuscript, among the SDSS pipeline 
classified low-redshift quasars ($z~<~0.3$) \citep{rg02,ro12}, \obj~ is collected as the target, because 
the \obj~ is the second mis-classified broad line AGN as well discussed in the following sections, and 
further due to its clean line profiles of [O~{\sc iii}] doublet without extended components. The manuscript 
is organized as follows. Section 2 shows the properties of the spectroscopic emission lines by different 
model functions with different considerations. Section 3 shows the properties of \obj~ in the BPT 
diagram. Section 4 discusses the probable physical origin of the uniqe properties of the mis-classified 
broad line AGN \obj. Section 5 gives our final summaries and conclusions. And in the manuscript, the 
cosmological parameters of $H_{0}~=~70{\rm km\cdot s}^{-1}{\rm Mpc}^{-1}$, $\Omega_{\Lambda}~=~0.7$ 
and $\Omega_{m}~=~0.3$ have been adopted.

\begin{figure*}
\centering\includegraphics[width = \textwidth,height=0.33\textwidth]{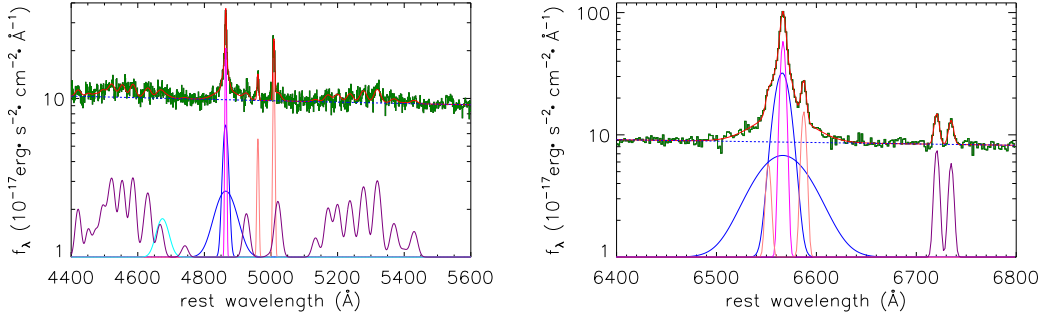}
\caption{The best fitting descriptions to the emission lines around H$\beta$ (left panel) and 
H$\alpha$ (right panel) by multiple Gaussian functions 
plus power-law continuum emissions. In each panel, solid dark green line shows the SDSS spectrum, solid red 
line shows the determined best-fitting results, dashed blue line shows the determined power law continuum 
emissions. In left panel, solid blue lines show the determined broad H$\beta$ described by two broad Gaussian 
functions, solid purple line shows the determined optical Fe~{\sc ii} emission features, solid cyan line shows 
the determined broad He~{\sc ii} line, solid magenta line shows the determined narrow H$\beta$, solid pink 
line shows the determined core components in [O~{\sc iii}] doublet. In right panel, solid blue lines show 
the determined two broad components in broad H$\alpha$ described by two broad Gaussian functions, solid magenta 
line shows the determined narrow H$\alpha$, solid pink lines show the determined [N~{\sc ii}] doublet, and 
solid purple lines show the determined [S~{\sc ii}] doublet. In order to show clear emission 
features, the Y-axis is in logarithmic coordinate in each panel.
}
\label{line}
\end{figure*}

\begin{figure*}
\centering\includegraphics[width = \textwidth,height=0.33\textwidth]{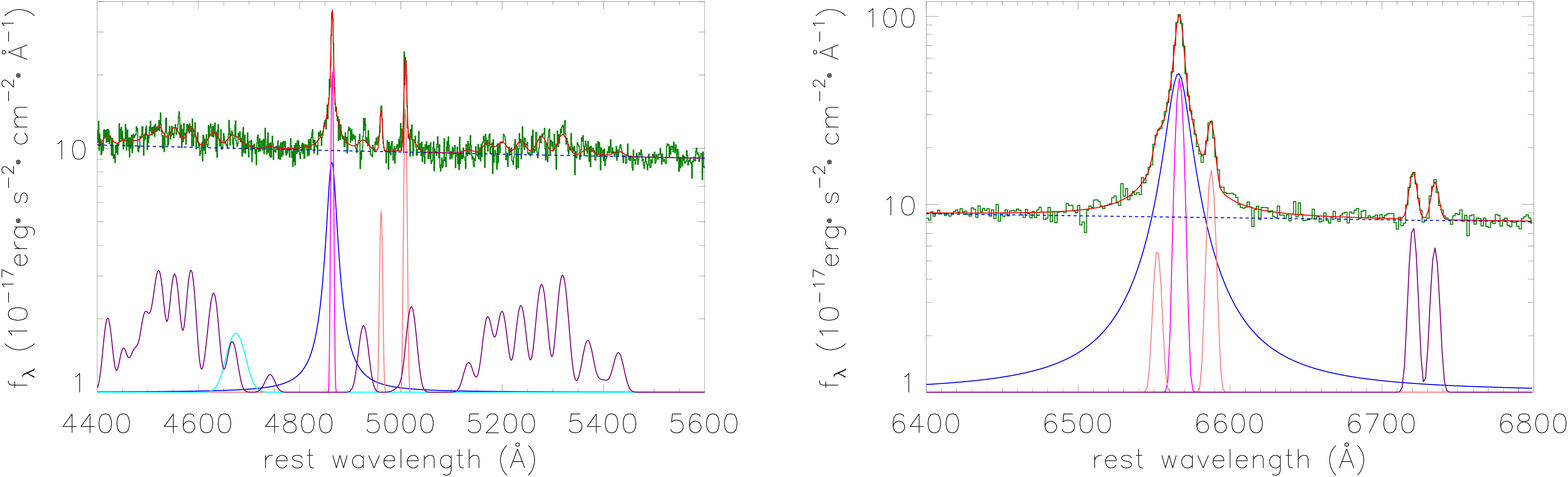}
\caption{Same as Fig.~\ref{line}, but the broad components in Balmer lines are described by Lorentz 
functions. In each panel, the solid blue line shows the Lorentz-function described broad Balmer line, 
the other line styles have the same meanings as those in Fig.~\ref{line}.
}
\label{llor}
\end{figure*}

\section{Properties of spectroscopic emission lines of \obj}

	Fig.~\ref{spec} shows the SDSS spectrum of \obj, with apparent broad emission lines indicating 
that \obj~ is undoubtfully a broad line AGN (type-1 AGN). In order to show the classification of \obj~ by 
flux ratios of narrow emission lines in the BPT diagram, the following emission line fitting procedures 
are applied to describe the emission lines of \obj, especially the emission lines of narrow H$\beta$, 
narrow H$\alpha$, [O~{\sc iii}]$\lambda4959,~5007$\AA~ doublet and [N~{\sc ii}]$\lambda6548,~6583$\AA 
doublet which will be applied in the following BPT diagram, within rest wavelength from 4400 to 
5600\AA\ and from 6400 to 6800\AA, which are fitted simultaneously by the following two different kinds 
of model functions. The fitting procedure is very similar as what we have done in \citet{zh16, zh17, 
zh21a, zh21m, zh21x}, and simply described as follows.

	For model A, Gaussian functions are applied to describe the emission lines as follows. Two narrow 
Gaussian functions are applied to describe the [O~{\sc iii}]$\lambda4959,~5007$\AA~ doublet. Here, as the 
following shown best-fitting results, there is no necessary to consider extended components of 
both [O~{\sc iii}]$\lambda4959,~5007$\AA~ doublet and the other narrow emission lines. Even two additional 
Gaussian components were applied to describe the probable extended components of [O~{\sc iii}]$\lambda4959,~5007$\AA~ 
doublet and the other narrow emission lines, the determined line fluxes of the extended components near to 
zero and smaller than determined uncertainties. Therefore, the [O~{\sc iii}]$\lambda4959,~5007$\AA~ doublet are 
clear in the \obj. Two narrow Gaussian functions are applied to describe the narrow H$\beta$ and narrow H$\alpha$. 
Two\footnote{More than two broad Gaussian functions have also been applied to describe the broad Balmer lines, 
however, the two or more broad Gaussian components are not necessary, because of the corresponding determined 
model parameters smaller than their uncertainties.} broad Gaussian functions are applied to describe the broad 
H$\beta$. Two broad Gaussian functions are applied to describe the broad H$\alpha$. Two\footnote{It is not necessary 
to consider extended components in [S~{\sc ii}] and [N~{\sc ii}] doublet. If additional Gaussian 
components are applied to describe probable extended components in the doublets, the determined line fluxes of the 
extended components near to zero and smaller than determined uncertainties. Therefore, in the manuscript, 
there are no considerations on extended components of the forbidden emission lines.} narrow Gaussian functions are applied to 
describe the [S~{\sc ii}]$\lambda6716,6731$\AA. One broad Gaussian function is applied to describe the 
broad He~{\sc ii} line. The broadened optical Fe~{\sc ii} template in \citet{kp10} is applied to describe 
the optical Fe~{\sc ii} emission features. One power law component is applied to describe the AGN continuum 
emissions underneath the emission lines around H$\beta$. One power law component is applied to describe 
the AGN continuum emissions underneath the emission lines around H$\alpha$. For the model parameters of 
the model functions in model A, the following restrictions are accepted. First, the flux of each Gaussian 
component is not smaller than zero. Second, the flux ratio of the [O~{\sc iii}] doublet ([N~{\sc ii}] 
doublet) is fixed to the theoretical value 3. Third, the Gaussian components of each forbidden line doublet 
have the same redshift and the same line width in velocity space. There are no further restrictions on the 
parameters of Balmer emission lines.

	For model B, the model functions are similar as the ones in model A, but the broad H$\beta$
(H$\alpha$) is described by one broad Lorentz function. And the same restrictions are accepted to
the model parameters in model B. The main objective to consider Lorentz function to describe the broad
Balmer lines is as follows. Not similar Gaussian function, Lorentz function always has sharp peak which
can lead to more smaller measured fluxes of narrow Balmer lines, which will have positive effects on the
classifications by flux ratios of narrow emission lines in BPT diagram, which will be discussed in the
following section.

	Through the Levenberg-Marquardt least-squares minimization technique, the emission lines around 
H$\beta$ and around H$\alpha$ can be well measured. The best fitting results are shown in Fig.~\ref{line} 
with $\chi^2/dof=1.18$ (summed squared residuals divided by degree of freedom) by model functions in 
model A, and in Fig~\ref{llor} with $\chi^2/dof=1.19$ by model functions in model B. The line parameters of 
each Gaussian component of narrow emission lines, Gaussian and Lorentz describe broad Balmer lines, and 
the power law continuum emissions are listed in Table~1.

\begin{table*}
\caption{Line parameters of the emission lines}
\begin{tabular}{lccl|ccl}
\hline\hline
name & $\lambda_0$ & $\sigma$ & flux &  $\lambda_0$ & $\sigma$ & flux  \\
\hline
&  \multicolumn{3}{c}{Gaussian Broad Balmer lines} & 
	\multicolumn{3}{c}{Lorentz Broad Balmer lines}  \\
&  \multicolumn{3}{c}{$\chi^2/dof=1.18$} &  \multicolumn{3}{c}{$\chi^2/dof=1.19$} \\
\hline
H$\beta_{B1}$ & 4863.91$\pm$0.78 & 30.29$\pm$4.65 & 121$\pm$16 &
	4863.54$\pm$0.42 & 20.28$\pm$2.22 & 247$\pm$11 \\
H$\beta_{B2}$ & 4863.57$\pm$0.09 & 7.51$\pm$0.88 & 109$\pm$13 &
        \dots & \dots & \dots \\
He~{\sc ii} & 4674.16$\pm$4.88 & 17.63$\pm$4.83 & 34$\pm$8 & 
	4673.97$\pm$4.78 & 17.04$\pm$4.71 & 32$\pm$8 \\
H$\beta_N$ & 4864.16$\pm$0.04 & 1.99$\pm$0.11 & 99$\pm$8 &
	4864.23$\pm$0.05 & 1.98$\pm$0.11 & 98$\pm$8 \\
\oiiin & 5008.69$\pm$0.18 & 2.48$\pm$0.15 & 85$\pm$6 &
	5008.73$\pm$0.18 & 2.44$\pm$0.15 & 83$\pm$6 \\
H$\alpha_{B1}$ & 6566.27$\pm$1.06 & 26.49$\pm$1.24 & 384$\pm$20 &
	6565.99$\pm$0.09 & 15.01$\pm$0.66 & 1150$\pm$20 \\
H$\alpha_{B2}$ & 6565.81$\pm$0.12 & 7.38$\pm$0.24 & 575$\pm$19 &
	\dots & \dots & \dots \\
H$\alpha_N$ & 6566.61$\pm$0.05 & 2.45$\pm$0.08 & 351$\pm$18 &
	6566.70$\pm$0.06 & 2.27$\pm$0.11 & 262$\pm$24 \\
\n6583 & 6587.33$\pm$0.12 & 2.35$\pm$0.12 & 86$\pm$5 &
        6587.53$\pm$0.12 & 2.28$\pm$0.11 & 81$\pm$4 \\
\sii & 6720.45$\pm$0.12 & 2.33$\pm$0.13 & 39$\pm$2 &
	6720.45$\pm$0.12 & 2.36$\pm$0.12 & 39$\pm$2 \\
\siii & 6734.83$\pm$0.12 & 2.34$\pm$0.13 & 28$\pm$2 &
	6734.83$\pm$0.13 & 2.36$\pm$0.13 & 29$\pm$2 \\
pow H$\beta$ & \multicolumn{3}{c}{$f_\lambda=(9.57\pm0.04)(\frac{\lambda}{5100\textsc{\AA}})^{-0.52\pm0.05}$} &
\multicolumn{3}{c}{$f_\lambda=(9.57\pm0.04)(\frac{\lambda}{5100\textsc{\AA}})^{-0.52\pm0.05}$} \\
pow H$\alpha$ & \multicolumn{3}{c}{$f_\lambda=(8.74\pm0.04)(\frac{\lambda}{6563\textsc{\AA}})^{-1.74\pm0.19}$} &
\multicolumn{3}{c}{$f_\lambda=(8.56\pm0.04)(\frac{\lambda}{6563\textsc{\AA}})^{-1.59\pm0.19}$} \\
\hline
\end{tabular}\\
{\bf Note:} The first column shows the information of emission component listed. 
[O~{\sc iii}]$\lambda5007$\AA(N) mean the narrow component of [O~{\sc iii}]$\lambda5007$\AA. H$\beta_N$ 
and H$\alpha_N$ mean the narrow H$\beta$ and the narrow H$\alpha$. The 'pow H$\beta$' and the 'pow H$\alpha$' 
mean the power law continuum emissions around H$\beta$ and around H$\alpha$, respectively. The second 
column to the fourth column show the line parameters of central wavelength in unit of \AA, second moment 
in unit of \AA~ and line flux in unit of $10^{-17}{\rm erg/\cdot~s^{-1}\cdot~cm^{-2}}$ of the determine 
components by Model A with two broad Gaussian functions (H$\beta_{B1}$, H$\beta_{B2}$ and H$\alpha_{B1}$, 
H$\alpha_{B2}$) applied to describe the broad Balmer lines. The fifth column to the seventh column show 
the line parameters of the determine components by Model B with one broad Lorentz function (H$\beta_{B1}$, 
H$\alpha_{B1}$) applied to describe the broad Balmer lines.
\end{table*}

	The different model functions clearly lead to quite different line parameters of narrow 
Balmer emission lines (especially line flux of narrow H$\alpha$) but similar line parameters of 
[O~{\sc iii}] and [N~{\sc ii}] doublets, which will lead to quite different flux ratios of [O~{\sc iii}] 
to narrow H$\beta$ (O3HB) and of [N~{\sc ii}] to narrow H$\alpha$ (N2HA).

\begin{figure*}
\centering\includegraphics[width = \textwidth,height=0.9\textwidth]{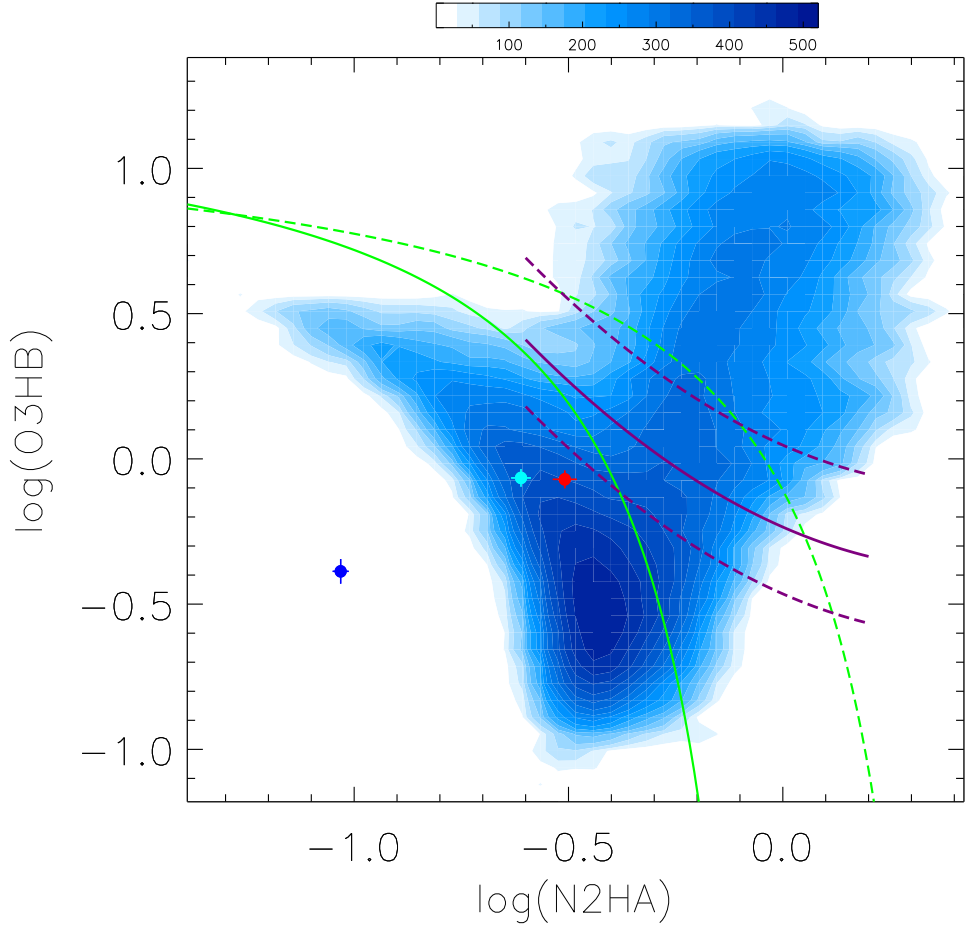}
\caption{The BPT diagram for more than 35000 narrow line objects (contour in bluish 
colors) and the mis-classified broad line AGN \obj~ (solid circles in red, cyan, purple) by O3HB versus 
N2HA. Solid and dashed green lines show the dividing lines reported in \citet{ka03} and in \citet{kb01} 
between HII galaxies, composite galaxies and AGN. Solid purple line and dashed purple lines show the dividing 
line between HII galaxies and AGN and the area for composite galaxies determined in our recent work 
in \citet{zh20} to determine the dividing lines in the BPT diagram through the powerful t-SNE technique.  
The contour is created by emission line properties of more than 35000 narrow emission-line 
galaxies discussed in \citet{zh20} collected from SDSS DR15. Corresponding number densities to different 
colors are shown in the color bar. Solid circles in blue, cyan and red represent the results of 
[$N2HA_{LA}$, $O3HB_{LA}$], [$N2HA_{UA}$, $O3HB_{UA}$] and [$N2HA_B$, $O3HB_{B}$], respectively.
}
\label{bpt}
\end{figure*}

\section{\obj\ in the BPT diagram}

\subsection{Flux ratios of narrow emission lines based on the model A}

	Based on the measured line parameters listed in Table~1 by model A, the H$\beta_{B1}$ and 
H$\alpha_{B1}$ are certainly from central BLRs, because of their line widths (second moment) quite 
larger than $900{\rm km\cdot~s^{-1}}$. The forbidden emission lines are certainly from central NLRs. 
Comparing with line widths of the [O~{\sc iii}] components, the determined narrow components of Balmer 
emissions, H$\beta_N$, and H$\alpha_N$, can be well accepted to come from central NLRs, because 
their line widths are smaller than the line width of the of [O~{\sc iii}] line.

	Besides the emission components discussed above, the determined broad component H$\beta_{B2}$ 
and H$\alpha_{B2}$ have line width larger than the line width of [O~{\sc iii}] component but smaller 
than $900{\rm km\cdot~s^{-1}}$, it is hard to confirm the broad components of H$\beta_{B2}$ and 
H$\alpha_{B2}$ with line width about $400{\rm km\cdot~s^{-1}}$ are from central BLRs.

	Finally, for the determined components shown in Fig.~\ref{line} and with parameters listed
in the second column to the fourth column in Table~1, the [N~{\sc ii}] and [O~{\sc iii}] doublets and narrow 
Balmer lines are from central NLRs. Therefore, considering the H$\beta_{B2}$ and H$\alpha_{B2}$ coming 
from central NLRs, the lower limit of flux ratio of [O~{\sc iii}]$\lambda5007$\AA~ to narrow H$\beta$ (
including two components of H$\beta_N$ and H$\beta_{B2}$) and lower limit of flux ratio of 
[N~{\sc ii}]$\lambda6583$\AA~ to narrow H$\alpha$ (including two components of H$\alpha_N$ and 
H$\alpha_{B2}$) can be estimated as
\begin{equation}
\begin{split}
O3HB_{LA}~=~\frac{f_{[O~\textsc{iii}]\lambda5007\textsc{\AA}(N)}}
	{f_{H\beta_N}~+~f_{H\beta_{B2}}}~=~ 0.41\pm0.04;\ \ \ \ \ \ \ \
N2HA_{LA}~=~\frac{f_{[N~\textsc{ii}]\lambda6583\textsc{\AA})}}
	{f_{H\alpha_N}~+~f_{H\alpha_{B2}}}~=~ 0.093\pm0.004
\end{split}
\end{equation}
If considering the H$\beta_{B2}$ and H$\alpha_{B2}$ coming from central BLRs, the upper limit of flux 
ratio of [O~{\sc iii}]$\lambda5007$\AA~ to narrow H$\beta$ (only one component H$\beta_N$) and the upper 
limit of flux ratio of [N~{\sc ii}]$\lambda6583$\AA~ to narrow H$\alpha$ (only one component H$\alpha_N$) 
can be estimated as
\begin{equation}
\begin{split}
O3HB_{UA}~=~\frac{f_{[O~\textsc{iii}]\lambda5007\textsc{\AA}(N)}}
	{f_{H\beta_N}}~=~ 0.86\pm0.06;\ \ \ \ \ \ \ \
N2HA_{UA}~=~\frac{f_{[N~\textsc{ii}]\lambda6583\textsc{\AA})}}
	{f_{H\alpha_N}}~=~ 0.245\pm0.013
\end{split}
\end{equation}

	Based on the determined narrow emission line ratios by model A, the \obj~ can be plotted
in the BPT diagram of O3HB versus N2HA in Fig.~\ref{bpt}. Considering the dividing lines in the BPT 
diagram as well discussed in \citet{ka03, kb06, kb19, zh20}, either [$N2HA_{UA}$,~$O3HB_{UA}$] or 
[$N2HA_{LA}$,~$O3HB_{LA}$] applied in the BPT diagram, the \obj~ can be well classified as a HII 
galaxy with few contributions of central AGN activities, through the properties of narrow emission lines.

\subsection{Flux ratios of narrow emission lines based on the model B}

	For the results by model B, the determined H$\beta_{B1}$ and H$\alpha_{B1}$ have line widths 
quite larger than $900{\rm km\cdot s^{-1}}$, therefore, the determined H$\beta_{B1}$ and H$\alpha_{B1}$ 
can be safely accepted to be from central BLRs. Certainly, the forbidden narrow lines are considered 
and well accepted from the central NLRs. Then, comparing with the line width $150{\rm km\cdot~s^{-1}}$ 
of [O~{\sc iii}]$\lambda$5007\AA, the determined components of narrow emission lines with line widths 
smaller than $150{\rm km\cdot~s^{-1}}$ can be safely accepted to be from central NLRs.

	Finally, for the determined components shown in Fig.~\ref{llor} and with parameters listed
in the fifth column to the seventh column in Table~1, the broad components H$\beta_{B1}$, H$\alpha_{B1}$
are truly from central BLRs, the [N~{\sc ii}] and [O~{\sc iii}] doublets and narrow Balmer lines are from 
central NLRs. Therefore, flux ratio of [O~{\sc iii}]$\lambda5007$\AA~ to narrow H$\beta$ can be estimated as
\begin{equation}
\begin{split}
O3HB_{B}~=~\frac{f_{[O~\textsc{iii}]\lambda5007\textsc{\AA}(N)}}
	{f_{H\beta_N}}~=~ 0.85\pm0.06;\ \ \ \ \ \ \ \ 
N2HA_{B}~=~\frac{f_{[N~\textsc{ii}]\lambda6583\textsc{\AA})}}
	{f_{H\alpha_N}}~=~ 0.31\pm0.02
\end{split}
\end{equation}

	Based on the model B, the \obj~ can be re-plotted in the BPT diagram of O3HB versus N2HA in 
Fig.~\ref{bpt}. The \obj~ can be well classified as a HII galaxy with few contributions of central AGN 
activities, through the properties of narrow emission lines.

	Finally, based on different model functions to describe emission lines and based on different 
considerations of emission components from central NLRs, the \obj~ can be well classified as a HII 
galaxy with few contributions of central AGN activities. In the manuscript, the \obj~ can be called 
as a mis-classified broad line AGN, based on the applications of BPT diagram.

\section{Physical origin of the mis-classified broad line AGN \obj?}

	In order to explain the mis-classified broad line AGN \obj, two reasonable methods can be 
mainly considered in the section, as what we have discussed in SDSS J1451+2709\ in \citet{zh21x}. 
On the one hand, there is one mechanism leading to stronger narrow Balmer emissions, such as the 
strong starforming contributions. On the other hand, there is one mechanism leading to weaker 
forbidden emission lines, such as the expected high electron densities in central NLRs.

\subsection{Strong Starforming contributions?}

	In the subsection, starforming contributions are checked, in order to explain the quite small 
value of O3HB, because of stronger starforming contributions leading to stronger narrow Balmer emission 
lines. In other words, there are two kinds of flux components included in the narrow Balmer lines and 
[O~{\sc iii}]$\lambda5007$\AA~ and [N~{\sc ii}]$\lambda6583$\AA, one kind of flux depending on central 
AGN activities: $f_{[O~\textsc{iii}]}(AGN)$, $f_{[N~\textsc{ii}]}(AGN)$, $f_{H\alpha}(AGN)$ and 
$f_{H\beta}(AGN)$, the other kind of flux depending on starforming: $f_{[O~\textsc{iii}]}(SF)$, 
$f_{[N~\textsc{ii}]}(SF)$, $f_{H\alpha}(SF)$ and $f_{H\beta}(SF)$. Then, the measured flux ratio 
$O3HB$ and $N2HA$, and the flux rations $O3HB(AGN)$ and $N2HA(AGN)$ depending on central AGN activities, 
and the flux ratios $O3HB(SF)$ and $N2HA(SF)$ depending on starforming, can be described as
\begin{equation}
\begin{split}
&O3HB~=~\frac{f_{[O~\textsc{iii}]}(AGN)~+~f_{[O~\textsc{iii}]}(SF)}{f_{H\beta}(AGN)~+~f_{H\beta}(SF)} \hspace{2cm}
	N2HA~=~\frac{f_{[N~\textsc{ii}]}(AGN)~+~f_{[N~\textsc{ii}]}(SF)}{f_{H\alpha}(AGN)~+~f_{H\alpha}(SF)} \\
&O3HB(AGN)~=~\frac{f_{[O~\textsc{iii}]}(AGN)}{f_{H\beta}(AGN)} \hspace{3cm} 
	O3HB(SF)~=~\frac{f_{[O~\textsc{iii}]}(SF)}{f_{H\beta}(SF)} \\
&N2HA(AGN)~=~\frac{f_{[N~\textsc{ii}]}(AGN)}{f_{H\alpha}(AGN)} \hspace{3cm} 
	N2HA(SF)~=~\frac{f_{[N~\textsc{ii}]}(SF)}{f_{H\alpha}(SF)} \\
&f_{[O~\textsc{iii}]}~=~f_{[O~\textsc{iii}]}(AGN)~+~f_{[O~\textsc{iii}]}(SF) \hspace{2cm} 
	f_{[N~\textsc{ii}]}~=~f_{[N~\textsc{ii}]}(AGN)~+~f_{[N~\textsc{ii}]}(SF) \\
& f_{H\beta}~=~f_{H\beta}(AGN)~+~f_{H\beta}(SF) \hspace{3cm} 
	f_{H\alpha}~=~f_{H\alpha}(AGN)~+~f_{H\alpha}(SF)
\end{split}
\end{equation}
where $f_{[O~\textsc{iii}]}$, $f_{[N~\textsc{ii}]}$, $f_{H\alpha}$ and $f_{H\beta}$ mean the measured
total line fluxes of [O~{\sc iii}]$\lambda5007$\AA, [N~{\sc ii}]$\lambda6583$\AA~ and narrow Balmer lines.

	Based on the three measured data points shown in Fig.~\ref{bpt} and the corresponding measured 
total line fluxes $f_{[O~\textsc{iii}]}$, $f_{[N~\textsc{ii}]}$, $f_{H\alpha}$ and $f_{H\beta}$ listed in 
Table~1 and discussed in Section 3, expected properties of starforming contributions 
$R_{SF}~=~f_{H\alpha}(SF)/f_{H\alpha}$ can be simply determined, through the following limitations. First, 
the determined flux ratios of $O3HB(AGN)$ and $N2HA(AGN)$ clearly lead the data points classified as AGN 
in the BPT diagram, the data points lying above the dividing line shown as solid green line in Fig.~\ref{bpt}. 
Second, the determined flux ratios of $O3HB(SF)$ and $N2HA(SF)$ clearly lead the data points classified 
as AGN in the BPT diagram, the data points lying below the dividing line shown as solid green line in 
Fig.~\ref{bpt}. Third, the ratios of $f_{H\alpha}(SF)$ to $f_{H\alpha}(AGN)$ are similar as the ratios 
of $f_{H\beta}(SF)$ to $f_{H\beta}(AGN)$.

\begin{figure*}
\centering\includegraphics[width = \textwidth,height=0.6\textwidth]{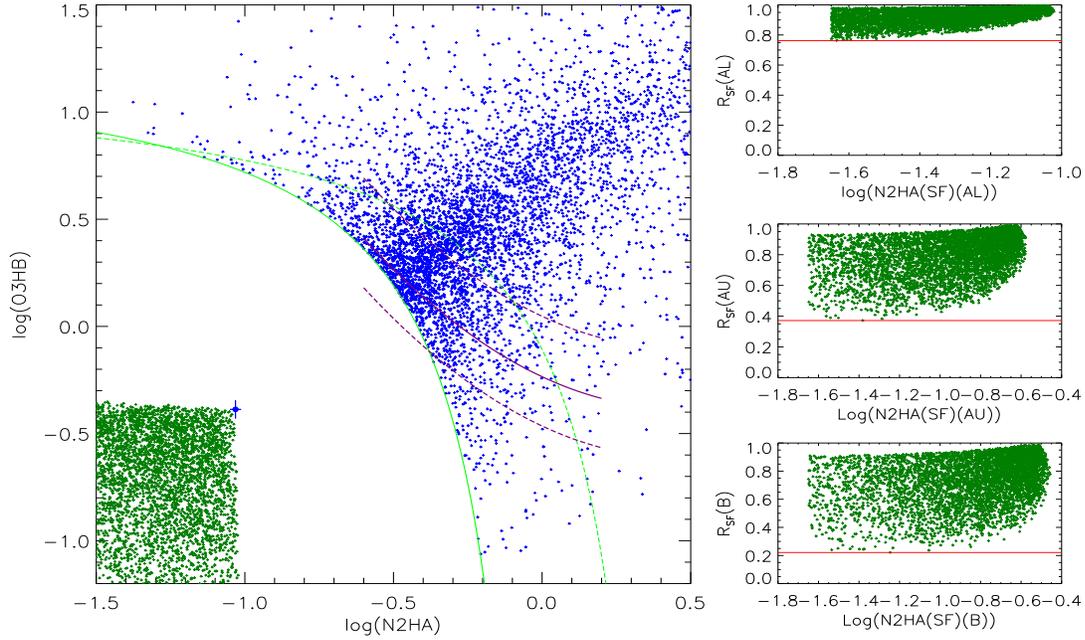}
\caption{Left panel shows the simulated 5000 couple data points of [$O3HB(AGN)$,~$N2HA(AGN)$] classified 
as AGN shown as blue pluses, and the simulated 5000 couple data points of [$O3HB(SF)$,~$N2HA(SF)$] classified 
as HII shown as dark green pluses, in the BPT diagram of O3HB versus N2HA, based on the narrow line fluxes 
determined by model A with considering H$\beta_{B2}$ and H$\alpha_{B2}$ from central NLRs. The solid blue 
circle plus error bars and the line styles are the same as those shown in Fig.~\ref{bpt}. Top right panel 
shows the dependence of $R_{SF}(AL)$ on $N2HA(SF)(AL)$, based on the narrow line fluxes determined by model 
A with considering H$\beta_{B2}$ and H$\alpha_{B2}$ from central NLRs. Middle right panel shows the 
dependence of $R_{SF}(AU)$ on $N2HA(SF)(AU)$, based on the narrow line fluxes determined by model A with 
considering H$\beta_{B2}$ and H$\alpha_{B2}$ from central BLRs. Bottom right panel shows the dependence of 
$R_{SF}(B)$ on $N2HA(SF)(B)$, based on the narrow line fluxes determined by model B. In each right panel, 
horizontal red line shows the position of the minimum value of $R_{SF}$.
}
\label{cbpt}
\end{figure*}

	Based on the model A with considering H$\beta_{B2}$ and H$\alpha_{B2}$ from NLRs, the narrow 
emission line fluxes are about $f_{[O~\textsc{iii}]}~\sim~85$, 
$f_{[N~\textsc{ii}]}~\sim~86$, $f_{H\alpha}~\sim~926$ and $f_{H\beta}~\sim~208$ in the units of 
$10^{-17}{\rm erg\cdot~s^{-1}\cdot~cm^{-2}}$. Model simulating results can be simply done by the 
following three steps. First and foremost, based on the measured values of $f_{[O~\textsc{iii}]}$, 
$f_{[N~\textsc{ii}]}$, $f_{H\alpha}$ and $f_{H\beta}$, value of $f_{[O~\textsc{iii}]}(AGN)$ can be 
randomly selected from 0 to 85, leading to the fixed 
$f_{[O~\textsc{iii}]}(SF)~=~85~-~f_{[O~\textsc{iii}]}(AGN)$; value of $f_{[N~\textsc{ii}]}(AGN)$ can 
be randomly selected from 0 to 86, leading to the fixed  
$f_{[N~\textsc{ii}]}(SF)~=~86~-~f_{[N~\textsc{ii}]}(AGN)$; value of $f_{H\alpha}(AGN)$ can be randomly 
selected from 0 to 926, leading to the fixed $f_{H\alpha}(SF)~=~926~-~f_{H\alpha}(AGN)$ 
and leading to the values of $f_{H\beta}(AGN)$ and $f_{H\beta}(SF)$ determined by
\begin{equation}
\begin{split}
&f_{H\beta}(AGN)~+~f_{H\beta}(SF)~=~208 \\
&f_{H\beta}(AGN)~=~\frac{f_{H\alpha}(AGN)}{f_{H\alpha}(SF)}~\times~f_{H\beta}(SF)
\end{split}
\end{equation}
Besides, based on the randomly selected values of $f_{[O~\textsc{iii}]}(AGN)$, $f_{[N~\textsc{ii}]}(AGN)$,
$f_{H\alpha}(AGN)$ and $f_{H\beta}(AGN)$, to determine whether the ratios of
$O3HB_s~=~\frac{f_{[O~\textsc{iii}]}(AGN)}{f_{H\beta}(AGN)}$ and
$N2HA_s~=~\frac{f_{[N~\textsc{ii}]}(AGN)}{f_{H\alpha}(AGN)}$ can lead to the data point [$O3HB_s$, $N2HA_s$]
lying above the dividing line shown as solid green line in the BPT diagram of O3HB versus N2HA. Only if the
data point [$O3HB_s$, $N2HA_s$] lies in the AGN region in the BPT diagram, the randomly selected values of
$f_{[O~\textsc{iii}]}(AGN)$, $f_{[N~\textsc{ii}]}(AGN)$, $f_{H\alpha}(AGN)$ are appropriate values. Therefore,
the randomly selected values each time are not always appropriate values. Last but not the least, after 
the first and the second steps are repeated tens of thousands of times, 5000 appropriate values can be 
collected for the $f_{[O~\textsc{iii}]}(AGN)$, $f_{[N~\textsc{ii}]}(AGN)$, and $f_{H\alpha}(AGN)$. Then, based on the 
results through the model A with considering H$\beta_{B2}$ and H$\alpha_{B2}$ from NLRs, among 66000 randomly 
selected values of $f_{[O~\textsc{iii}]}(AGN)$ from 0 to 85, of $f_{[N~\textsc{ii}]}(AGN)$ from 0 to 86 and of 
$f_{H\alpha}(AGN)$ from 0 to 926, there are 5000 couple data points of [$O3HB(AGN)$,~$N2HA(AGN)$] 
classified as AGN, and corresponding 5000 couple data points of [$O3HB(SF)$,~$N2HA(SF)$] classified as HII, 
in the BPT diagram of O3HB versus N2HA, shown in the left panel of Fig.~\ref{cbpt}. And top right panel of 
Fig.~\ref{cbpt} shows the dependence of $R_{SF}(AL)$ on $N2HA(SF)(AL)$, with the determined minimum value 
76\% of the $R_{SF}(AL)$.

\begin{figure*}
\centering\includegraphics[width = \textwidth,height=0.333\textwidth]{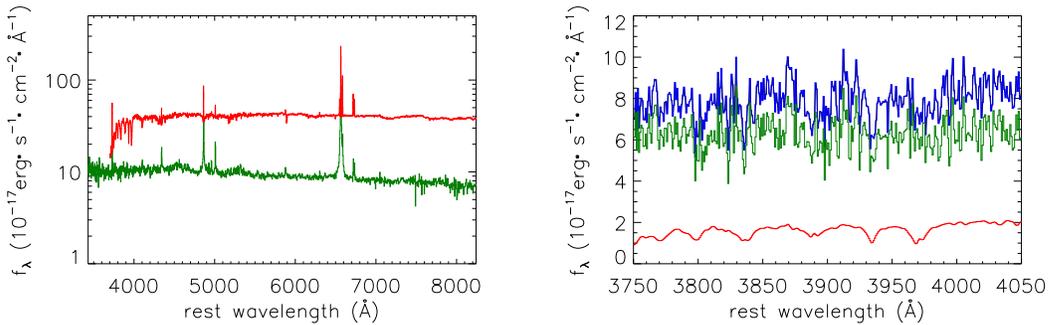}
\caption{Left panel shows the SDSS spectrum of \obj~ in dark green and the mean spectrum of 
HII galaxies in red. Right panel shows the composite spectrum around 4000\AA, including 20\% 
starforming contributions. In right panel, solid blue line shows the composite spectrum, solid 
dark green line shows the SDSS spectrum, and solid red line shows the starforming contributions.
}
\label{sp2}
\end{figure*}

	Similarly, based on the model A with considering H$\beta_{B2}$ and H$\alpha_{B2}$ from BLRs, 
the narrow emission line fluxes are about $f_{[O~\textsc{iii}]}~\sim~85$, 
$f_{[N~\textsc{ii}]}~\sim~86$, $f_{H\alpha}~\sim~351$ and $f_{H\beta}~\sim~99$ in the units of 
$10^{-17}{\rm erg\cdot~s^{-1}\cdot~cm^{-2}}$. Then, among 22000 randomly selected values of 
$f_{[O~\textsc{iii}]}(AGN)$, of $f_{[N~\textsc{ii}]}(AGN)$ and of $f_{H\alpha}(AGN)$, there are 5000 
couple data points of [$O3HB(AGN)$,~$N2HA(AGN)$] classified as AGN, and corresponding 5000 couple data 
points of [$O3HB(SF)$,~$N2HA(SF)$] classified as HII, in the BPT diagram of O3HB versus N2HA. Here, we 
do not show the results in the BPT diagram, which are totally similar as the those shown in the left 
panel of Fig.~\ref{cbpt}. And the middle right panel of Fig.~\ref{cbpt} shows the dependence of 
$R_{SF}(AU)$ on $N2HA(SF)(AU)$, with the determined minimum value 37\% of the $R_{SF}(AU)$.

	Based on the model B, the narrow emission line fluxes are about  
$f_{[O~\textsc{iii}]}~\sim~83$, $f_{[N~\textsc{ii}]}~\sim~81$, $f_{H\alpha}~\sim~262$ and 
$f_{H\beta}~\sim~98$ in the units of $10^{-17}{\rm erg\cdot~s^{-1}\cdot~cm^{-2}}$. Then, among 15000
randomly selected values of $f_{[O~\textsc{iii}]}(AGN)$, of $f_{[N~\textsc{ii}]}(AGN)$ and of 
$f_{H\alpha}(AGN)$, there are 5000 couple data points of [$O3HB(AGN)$,~$N2HA(AGN)$] classified as AGN, 
and corresponding 5000 couple data points of [$O3HB(SF)$,~$N2HA(SF)$] classified as HII, in the BPT 
diagram of O3HB versus N2HA. Here, we do not show the results in the BPT diagram, which are totally 
similar as the those shown in the left panel of Fig.~\ref{cbpt}. And the bottom right panel of 
Fig.~\ref{cbpt} shows the dependence of $R_{SF}(B)$ on $N2HA(SF)(B)$, with the determined minimum 
value 22\% of the $R_{SF}(B)$.

	Therefore, the determined starforming contributions $R_{SF}$ should be larger than 22\%. The 
strong starforming contributions indicate there should be apparent absorption features from host 
galaxy in the spectrum of \obj. Fig.~\ref{sp2} shows one composite spectrum created by 0.8 times of the
SDSS spectrum of \obj~ plus a mean HII galaxy with continuum intensity at 5100\AA~ about 0.2 times
of the continuum intensity at 5100\AA~ of the SDSS spectrum of \obj. Here, the mean spectrum of HII
galaxies are created by the large sample of 1298 HII galaxies with signal-to-noise larger than 30\ 
in SDSS DR12. There should be absorption features around 4000\AA, clearly indicating that the 
starforming contributions could be well applied to explain the unique properties of the mis-classified 
broad line AGN \obj.

	Before the end of the subsection, there is one point we should note. As discussed in \citet{kh09} 
for galaxies around the boundary as defined in \citet{ka03}, contributions to [O~{\sc iii}] emissions 
by star-formations are predicted to be typically 40 to 50\%, which are roughly agreement with our 
determined $R_{SF}$ with minimum value of about 22\% and with maximum value of about 76\%, providing 
further clues to support the starforming contributions to explain the unique properties of the 
mis-classified broad line AGN \obj. However, one probable question is proposed why we did not see 
apparent contribution of AGN activities to the narrow emission lines in \obj? Actually, as the results 
shown in Fig.~\ref{cbpt} for the simulating results, the separated appropriate contributions of AGN 
activities to narrow emission lines, $f_{[O~\textsc{iii}]}(AGN)$, $f_{[N~\textsc{ii}]}(AGN)$, 
$f_{H\alpha}(AGN)$, are randomly determined and lead to the data points on AGN activities apparently 
lying in the AGN region in the BPT diagram. Therefore, the \obj~ including AGN activities but lying in 
the HII region in the BPT diagram is mainly due to mixed contributions of star-formations and AGN activities, if 
considering starforming contributions as the preferred model to explain the unique properties of the 
mis-classified broad line AGN \obj. Moreover, as the shown results in Fig.~\ref{cbpt}, there could be 
dozens of broad line quasars (a sample of tens of mis-classified quasars will be reported and discussed 
in one of our being prepared manuscripts) lying in the HII regions in the BPT diagrams with applications 
of narrow emission line properties. Certainly, intrinsic physical origin of the mis-classification in 
BPT diagram is still uncertain, further efforts are necessary.

\subsection{Compressed central NLRs?}

	Besides the starforming contributions, high electron density in NLRs can be also applied to 
explain the unique properties of the mis-classified broad line AGN \obj, because the high electron 
density near to the critical electron densities of the forbidden emission lines can lead to suppressed 
line intensities of forbidden emission lines but positive effects on strengthened Balmer emission lines.

	It is not hard to determine electron density in NLRs, such as through the flux ratios of 
[S~{\sc ii}]$\lambda6717,6731$\AA~ doublet as well discussed in \citet{po14, sa16, kn19}. Based on the 
measured line fluxes of [S~{\sc ii}] doublet listed in Table~1, the flux ratio of 
[S~{\sc ii}]$\lambda6716$\AA~ to [S~{\sc ii}]$\lambda6731$\AA~ lead the electron density to be 
estimated around ${\rm 200cm^{-3}}$, a quite normal value, quite smaller than the critical densities 
around ${\rm 10^5 cm^{-3}}$ to [O~{\sc iii}] and [N~{\sc ii}] doublet.

\section{Summaries and Conclusions}

Finally, we give our summaries and conclusions as follows.
\begin{itemize}
\item Emission lines of the blue quasar \obj~ can be well measured by two different models, broad Balmer 
	lines described by mode A with broad Gaussian functions and by model B with broad Lorentz functions, 
	leading to different flux ratios of narrow emission lines.
\item Different flux ratios of narrow emission lines determined by different model functions and with different 
	considerations, the \obj~ can be well classified as a HII galaxy in the BPT diagram (a mis-classified 
	broad line AGN), although the \obj~ actually is a normal broad line AGN.
\item Two reasonable methods are proposed to explain the unique properties of the mis-classified broad line AGN 
	\obj, strong starforming contributions leading to stronger narrow Balmer emissions, and compressed NLRs with
	high electron densities leading to suppressed forbidden emissions.
\item Once considering the starforming contributions, at least 20\% starforming contributions should be preferred 
	in the mis-classified broad line AGN \obj, which will lead to apparent absorption features around 
	4000\AA, indicating strong starforming contributions should be preferred in the mis-classified broad line 
	AGN \obj.
\item Once considering the compressed NLRs with high electron densities, the expected electron densities should 
	be around ${\rm 10^5cm^{-3}}$. However, the estimated electron density is only around ${\rm 200cm^{-3}}$ based on
	the flux ratio of [S~{\sc ii}]$\lambda6716$\AA~ to [S~{\sc ii}]$\lambda6731$\AA. Therefore, the compressed
	NLRs with high electron densities are not preferred in the mis-classified broad line AGN \obj.
\item The reported second mis-classified broad line AGN \obj~ strongly indicate that there should be extremely unique 
	properties of \obj~ which are currently not detected, or indicate that there should be a small sample of 
	mis-classified broad line AGN similar as the \obj.
\item Narrow emission line properties should be carefully determined in Type-1 AGN.
\end{itemize}

\begin{acknowledgements}
We gratefully acknowledge the anonymous referee for giving us constructive comments and 
suggestions to greatly improve our paper. Zhang X. G. gratefully acknowledges the kind support of Starting 
Research Fund of Nanjing Normal University, and the kind support of NSFC-12173020. Cao Y., Zhao S. D., 
Zhu X. Y., Yu H. C. and Wang Y. W. gratefully acknowledge the kind support of DaChuang project of NanJing 
Normal University for Undergraduate students. This manuscript has made use of the data from the SDSS 
projects. The SDSS-III web site is http://www.sdss3.org/. SDSS-III is managed by the Astrophysical 
Research Consortium for the Participating Institutions of the SDSS-III Collaborations. 
\end{acknowledgements}

\label{lastpage}

\end{document}